\documentclass[twocolumn,showpacs,preprintnumbers,amsmath,amssymb,eqsecnum]{revtex4}
\usepackage{graphicx}
\begin{document}
\preprint{IMAFF-RCA-03-04}
\title{You need not be afraid of phantom energy}

\author{Pedro F. Gonz\'{a}lez-D\'{\i}az}
\affiliation{Centro de F\'{\i}sica ``Miguel A. Catal\'{a}n'', Instituto de
Matem\'{a}ticas y F\'{\i}sica Fundamental,\\ Consejo Superior de
Investigaciones Cient\'{\i}ficas, Serrano 121, 28006 Madrid (SPAIN).}
\date{\today}
\begin{abstract}
Phantom energy which violates the dominant-energy condition and is
not excluded by current constraints on the equation of state may
be dominating the evolution of the universe now. It has been
pointed out that in such a case the fate of the universe may be a
big rid where the expansion is so violent that all galaxies,
planet and even atomic nuclei will be successively ripped apart in
finite time. Here we show however that there are certain unified
models for dark energy which are stable to perturbations in matter
density where the presence of phantom energy does not lead to such
a cosmic doomsday.
\end{abstract}


\maketitle

\section{Introduction}
WMAP [1] has confirmed it with the highest accuracy: Nearly
seventy percent of the energy in the universe is in the form of
dark energy -possibly one of most astonishing discoveries ever
made in science. Moreover, recent observations do not exclude, but
actually suggest a value even smaller than -1 for the parameter of
the equation of state, $\omega$, characterizing that dark energy
[2]. That means that for at least a perfect-fluid equation of
state, the absolute value for negative pressure exceeds that for
positive energy density, i.e. $\rho+p<0$, and hence it follows
that the involved violation of the dominant-energy condition might
allow the existence of astrophysical or cosmological wormholes. A
most striking consequence from dark energy with $\omega<-1$ has
been however pointed out [3]. It is that in a finite time the
universe will undergo a catastrophic "big rip". Big rip is a term
coined by Caldwell [3] that corresponds to a new cosmological
model in which the scale factor blows up in a finite time because
its cosmic acceleration is larger than that is induced by a
cosmological constant, making in this way every component of the
universe to go beyond the horizon of all other universe's
components in finite cosmic time. If dark energy is phantom
energy, i.e. if dark energy is characterized by an equation of
state with $\omega=p/\rho <-1$, and hence there is a violation of
the dominant-energy condition, $\rho+p<0$, then the phantom-energy
density is still positive though it will first increase from a
finite small initial value up to infinite in a finite time,
thereafter steadily decreasing down to zero as time goes to
infinity. A state with infinite energy density at finite
cosmological time is certainly an unusual state in cosmology. To
an observer on the Earth, this state coincides with the
above-mentioned big rip where the universe dies after ripping
successively apart all galaxies, our solar system, the Earth
itself, and finally molecules, atoms, nuclei and nucleons [3]. For
a general cosmological model with phantom energy, the time at
which that big rip would take place depends on both the initial
size of the universe and the value of $\omega$ in such a way that,
the larger the absolute value of these quantities, the nearer the
big rip will occur. The behaviour of the universe after the big
rip is in some respects even more bizarre than the big rip itself,
as its size then steadily decreases from infinite down to zero at
infinite time. In case that a generic perfect-fluid equation of
state, $p=\omega\rho$, with $\omega<-1$ is considered, the above
new behaviours show themselves immediately. For flat geometry, the
scale factor is then given by [4]:
\begin{equation}
a(t)\propto\left[e^{C_1(1+\omega)t}-C_2
e^{-C_1(1+\omega)t}\right]^{2/[3(1+\omega)]} ,
\end{equation}
with $C_1>0$ and $0<C_2<1$. We note that for $\omega<-1$, in fact
$a\rightarrow\infty$ as $t\rightarrow t_* =\ln
C_2/[C_1(1+\omega)]$. This marks the time at big rip and the onset
of the contracting phase for $t>t_*$.

This report aims at showing that the above-described emergence of
a cosmic doomsday at which the big rip occurred and the subsequent
unconventional evolution of the universe can both be avoided while
keeping the phantom energy condition, $\rho+p<0$, $\rho >0$,
$\omega <-1$, on the dark energy if, instead of a quintessential
description of dark energy based on an equation of state
$p=\omega\rho$, with $\omega<-1$, we consider a suitable
generalization of the Chaplygin-gas model which, at sufficiently
late times, does not show observable non-physical oscillations and
exponential blowup in the matter density perturbations [5] that
are present in current unstable Chaplygin-gas cosmic models [6].
The latter models describe a single substance which is
characterized by an equation of state [6] $p=-A\rho^{-\alpha}$,
where $A$ is a positive-definite constant and $\alpha$ is a
parameter which may take on any real positive values. This
equation of state has been shown to represent the stiff that
simultaneously describes dark matter and dark energy, but gives
rise to instabilities stemming from the unobserved existence of
oscillations and exponential blowup in the perturbation
power-spectrum which arise whenever the speed of sound in nonzero
[5].

The paper can be outlined as follows. In Sec. II we generalize the
cosmic Chaplygin-gas models in such a way that the resulting
models can be made stable and free from unphysical behaviours even
when the vacuum fluid satisfies the phantom energy condition. The
Friedmann equations for models which show and do not show
unphysical behaviours are solved in Sec. III, checking that in the
latter case the phantom energy condition does not imply any
emergence of a big rip in finite time. We finally conclude in Sec.
IV.

\section{Generalized cosmic Chaplygin-gas models}
We introduce here some generalizations from cosmic Chaplygin-gas
model that also contain an adjustable initial parameter $\omega$.
In particular, we shall consider a generalized gas whose equation
of state reduces to that of current Chaplygin unified models for
dark matter and energy in the limit $\omega\rightarrow 0$ and
satisfies the following conditions: (i) it becomes a de Sitter
fluid at late time and when $\omega=-1$, (ii) it reduces to
$p=\omega\rho$ in the limit that the Chaplygin parameter
$A\rightarrow 0$, (iii) it also reduces to the equation of state
of current Chaplygin unified dark matter models at high energy
density, and (iv) the evolution of density perturbations derived
from the chosen equation of state becomes free from the
above-mentioned pathological behaviour of the matter power
spectrum for physically reasonable values of the involved
parameters, at late time. We shall see that these generalizations
retain a big rip if they also show unphysical oscillations and
exponential blowup leading to instability (i.e. if they do not
satisfy condition (iv)), but if the latter effects are avoided
then the evolution of the scale factor recovers a rather
conventional pattern, without any big rip or contracting phase.

An equation of state that can be shown to satisfy all the above
conditions (i) - (iv) is
\begin{equation}
p=-\rho^{-\alpha}\left[C+\left(\rho^{1+\alpha}-
C\right)^{-\omega}\right] ,
\end{equation}
where
\begin{equation}
C=\frac{A}{1+\omega}-1 ,
\end{equation}
with $A$ a constant which now can take on both positive and
negative values, and $0>\omega>-\ell$, $\ell$ being a positive
definite constant which can take on values larger than unity. By
integrating the cosmic conservation law for energy we get for the
energy density
\begin{equation}
\rho(a)=\left[C +\left(1+\frac{B}{a^{3(1+\alpha)(1+
\omega)}}\right)^{\frac{1}{1+\omega}}\right]^{\frac{1}{1+\alpha}}
,
\end{equation}
where $B$ is a positive integration constant. Let us now define
the effective expressions of the state equation parameter and
speed of sound, which respectively are given by
\begin{equation}
\omega^{{\rm eff}}= \frac{p}{\rho}= -\frac{C+D(a)^{-1}}{C+D(a)}
\end{equation}
\begin{eqnarray}
&&c_s^{{\rm eff}2}=\frac{\partial p}{\partial\rho}=\nonumber\\ &&
\frac{\alpha
C\left(D(a)^{1+\alpha}-1\right)+(C+D(a))(\alpha+\omega(1+
\alpha))}{(C+D(a))D(a)^{1+\omega}}
\end{eqnarray}
with
\begin{equation}
D(a)=\left(1+\frac{B}{a^{3(1+\alpha)(1+
\omega)}}\right)^{\frac{1}{1+\omega}} .
\end{equation}
One can then interpret the model by taking the limit of these
parameters as $a\rightarrow 0$ and $a\rightarrow\infty$, at which
limits they respectively become $\omega^{{\rm eff}}\rightarrow 0$
and $c_s^{{\rm eff}2}\rightarrow 0$ (that correspond to the
pressureless CDM model), and $\omega^{{\rm eff}}\rightarrow -1$
and $c_s^{{\rm eff}2}\rightarrow\alpha+\omega(1+\alpha)$ (that
correspond to a pure cosmological constant). Such as we have
defined it so far, the present model does not satisfy condition
(iv) above, as the evolution of density perturbations $\delta_k$
with wave vector $k$ [5]
\begin{equation}
\delta_{k}''+F(\omega^{{\rm eff}},c_s^{{\rm eff}})\delta_{k}'
-G(\omega^{{\rm eff}},c_s^{{\rm eff}},k)\delta_k =0 ,
\end{equation}
where the prime denotes differentiation with respect to $\ln a$,
and
\begin{equation}
F(\omega^{{\rm eff}},c_s^{{\rm eff}}) =2+\xi-3(2\omega^{{\rm
eff}}-c_{s}^{{\rm eff}2})
\end{equation}
\begin{equation}
G(\omega^{{\rm eff}},c_s^{{\rm eff}},k)
=\frac{3}{2}\left(1-6c_{s}^{{\rm eff}2} +8\omega^{{\rm
eff}}-3\omega^{{\rm eff}2}\right)-\left(\frac{kc_{s}^{{\rm
eff}}}{aH}\right)^2 ,
\end{equation}
with $H$ the Hubble parameter,
\[\xi=-\frac{2}{3}\left[1+\left(\frac{1}{\Omega_M}-
1\right)a^{3(1+\alpha)(1+\omega)}\right]^{-1} \] and $\Omega_M$
the CDM density defined from Eq. (2.3) in the limit $a\rightarrow
0$, shows oscillations and exponential blowup because $c_s^{{\rm
eff}}$ is generally nonzero.

\section{Avoiding the big rip}
In case that $\omega<-1$ for the large values of the scale factor
for which dominance of the dark-phantom energy is expected, one
can approximate the Friedmann equation that corresponds to the
considered model as follows
\begin{equation}
\left(\frac{\dot{a}}{a}\right)^2 \simeq L^2 a^3,
\end{equation}
in which $L^2=8\pi GB^{-1/[(1+\alpha)(|\omega|-1)]}$. The solution
to this equation is
\begin{equation}
a(t)\simeq\left(a_0^{-3/2} -\frac{3L(t-t_0)}{2}\right)^{-2/3},
\end{equation}
where $a_0$ and $t_0$ are the initial values of the scale factor
and time, respectively. It is easy to see that in the considered
case the phantom energy condition always satisfies $p+\rho<0$ and
there will be a big rip, taking place now at a time
\begin{equation}
t_* \simeq\frac{2}{3a_0^{3/2}L} ,
\end{equation}
followed as well by a contracting phase where the size of the
universe vanishes as $t\rightarrow\infty$. Note that, as it was
pointed out before, the time at which the big rip occurs turns out
to depend on the initial size of the universe $a_0$, in such a way
that the big rip becomes nearer as $a_0$ is made larger. Thus, the
above model, which is actually excluded because it shares the same
kind of instabilities as the original cosmic Chaplygin-gas model
[5], shows a cosmological big rip.

In order to allow for both stability and compatibility with
observations, we consider next a model in which $c_{s}^{{\rm
eff}}\rightarrow 0$ as $t\rightarrow\infty$ and the nonzero value
of parameter $B$ is small enough. The first of these conditions
can be achieved by simply imposing $\alpha+\omega(1+\alpha)=0$,
i.e.
\begin{equation}
1+\alpha=\frac{1}{1+\omega} .
\end{equation}
The equation of state and the expression for the energy density
are then reduced to read:
\begin{equation}
p=-\rho^{-\alpha}\left[C+\left(\rho^{1+\alpha}-
C\right)^{\frac{\alpha}{1+\alpha}}\right] ,
\end{equation}
\begin{equation}
\rho(a)=\left[C
+\left(1+\frac{B}{a^{3}}\right)^{1+\alpha}\right]^{\frac{1}{1+\alpha}}
.
\end{equation}
This equation of state satisfies then all the condition (i) - (iv)
imposed above.

According to condition (3.4) the phantom-dark energy with
$\omega<-1$ immediately implies that $\alpha<-1$ too. In such a
case we can check that $p+\rho<0$ for all values of the scale
factor and, in order to ensure positiveness of the energy density,
we must have $A=-|A|$ and keep $B>0$. Then, for the large values
of the scale factor for which phantom-dark energy is expected to
dominate over matter, the energy density can be approximated to
\begin{equation}
\rho\simeq\left(\frac{|A|}{|\omega|-1}\right)^{-(|\omega|-1)}\left(1
+\frac{B}{|A|(|\omega|-1)a^3}\right) .
\end{equation}
We can now write for the Friedmann equation
\begin{equation}
\left(\frac{\dot{a}}{a}\right)^2\simeq
\tilde{A}\left(1+\frac{\tilde{B}}{a^3}\right) ,
\end{equation}
where
\begin{equation}
\tilde{A}=\ell_p^2\left(\frac{|\omega|-1}{|A|}\right)^{|\omega|-1}
> 0
\end{equation}
\begin{equation}
\tilde{B}=\frac{B}{(|\omega|-1)|A|} > 0 ,
\end{equation}
with $\ell_p^2=8\pi G/3$. The solution to Eq. (3.8) is
\begin{equation}
a(t)\simeq D\left(C_0 e^{-\frac{3}{2}\sqrt{\tilde{A}}(t-t_0)}
-e^{\frac{3}{2}\sqrt{\tilde{A}}(t-t_0)}\right)^{\frac{2}{3}} ,
\end{equation}
where
\begin{equation}
D=a_0\left(\frac{\mu}{4C_0}\right)^{1/3} ,
\end{equation}
\begin{equation}
C_0=\frac{\sqrt{1+\mu}-1}{\sqrt{1+\mu}+1}
\end{equation}
and
\begin{equation}
\mu=\tilde{B}a_0^{-3} .
\end{equation}
We notice that $a\rightarrow a_0$ as $t\rightarrow t_0$ and
$a\rightarrow\infty$ as $t\rightarrow\infty$, and hence there is
not a big rip for this solution.

\section{Conclusion}
It appears then that if we choose a general equation of state for
dark energy which is reasonably free from instabilities and
unphysical effects, then a phantom energy can be predicted which
does not show any big rip at finite time. The key difference
between the scale factor given by Eq. (3.11) and that given by Eq.
(1.1) is in the sign of the overall exponent of the
right-hand-side; while in Eq. (1.1) it is negative for
$\omega<-1$, in Eq. (3.11) it is positive for the same case. Thus,
cosmology can co-exist with these phantoms in a quite safe manner.

\acknowledgements

\noindent The author thanks Mariam Bouhmadi and Carmen L. Sig\"{u}enza
for useful conversations. This work was supported by MCYT under
Research Project No. BMF2002-03758.

\end{document}